\documentclass[a4paper]{jpconf}
\usepackage{graphicx}

\begin{document} 

\title{Nuclear medium effects from hadronic atoms} 

\author{E. Friedman, A. Gal} 

\address{Racah Institute of Physics, The Hebrew University, Jerusalem 91904, 
Israel} 

\ead{elifried@vms.huji.ac.il, avragal@vms.huji.ac.il} 

\begin{abstract} 
The state of the art in the study of 
$\pi^-$, $K^-$ and $\Sigma^-$ atoms, along with the in-medium nuclear 
interactions deduced for these hadrons, is reviewed. A special emphasis 
is placed on recent developments in $\bar K$--nuclear physics, where 
a strongly attractive density dependent $K^-$--nuclear potential of order 
150--200 MeV in nuclear matter emerges by fitting $K^-$--atom data. 
This has interesting repercussions on $\bar K$ quasibound {\it nuclear} 
states, on the composition of strange hadronic matter and on $\bar K$ 
condensation in self bound hadronic systems. 
\end{abstract}

\section{Introduction} 
\label{sec:intro} 
\vspace{0.5cm}

\begin{table}[b]  
\caption{Hadronic atom scenarios which except for $\bar p$ atoms are discussed 
in the present review.} 
\label{tab:scenarios} 
\begin{center} 
\begin{tabular}{llll} 
\br 
hadron&${\rm Re}~V_{\rm opt}$&${\rm Im}~V_{\rm opt}$&comments\\ 
\mr 
$\pi^-$&repulsive in bulk&moderate&excellent data (100)\\
       &attractive on surface&    &well understood\\ 
$\Sigma^-$&repulsive in bulk&moderate&limited data (23)\\ 
         &attractive outside&        &poorly understood\\ 
$K^-$&attractive&absorptive&good data (65)\\ 
     &{\it deep or shallow?}&  &{\it open problems}\\ 
$\bar p$&model dependent&very absorptive&excellent data (90)\\ 
\br 
\end{tabular} 
\end{center} 
\end{table} 

Hadronic atoms have played an important role in elucidating in-medium 
properties of hadron-nucleon ($hN$) interactions near threshold 
\cite{bfg97,fg07}. 
Hadronic atom data consist primarily of strong 
interaction level shifts, widths and yields, derived from $h^-$-atom X-ray 
transitions. These data are analyzed in terms of optical potentials 
$V_{\rm opt}^{h}=t_{hN}(\rho)\rho$ which are functionals of the nuclear 
density $\rho(r)$, capable of handling large data sets across the periodic 
table in order to identify characteristic entities that may provide a link 
between experiment and microscopic approaches. Here, $t_{hN}(\rho)$ is 
a density dependent in-medium $hN$ $t$ matrix at threshold, satisfying the 
low-density limit $t_{hN}(\rho) \to t_{hN}^{\rm free}$, with the free-space 
$t$ matrix $t_{hN}^{\rm free}$, as $\rho \to 0$. A schematic summary of 
lessons gained by analyzing the available hadronic atom data in terms of 
optical potentials is given in Table~\ref{tab:scenarios}, where the number 
of data points included in these analyses is shown in parentheses in the 
last column. A more exhaustive discussion of $\pi^-$, $\Sigma^-$ and $K^-$ 
atoms follows below. 

\section{Partial restoration of chiral symmetry from pionic atoms} 
\label{sec:pi} 
\vspace{0.5cm} 

\begin{figure}[hbt]  
\begin{center} 
\includegraphics[width=0.45\textwidth]{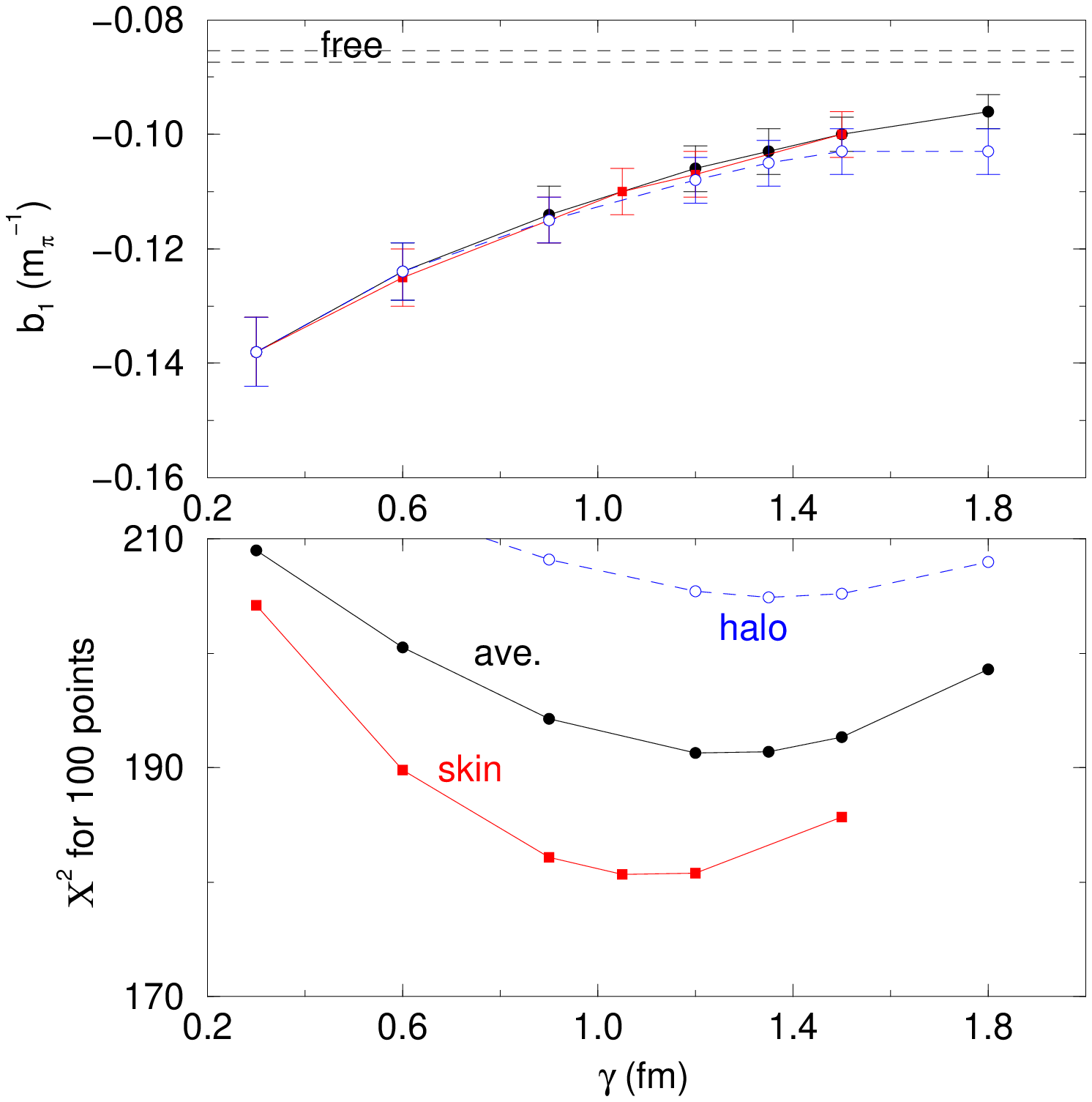} 
\hspace{3mm} 
\includegraphics[width=0.45\textwidth]{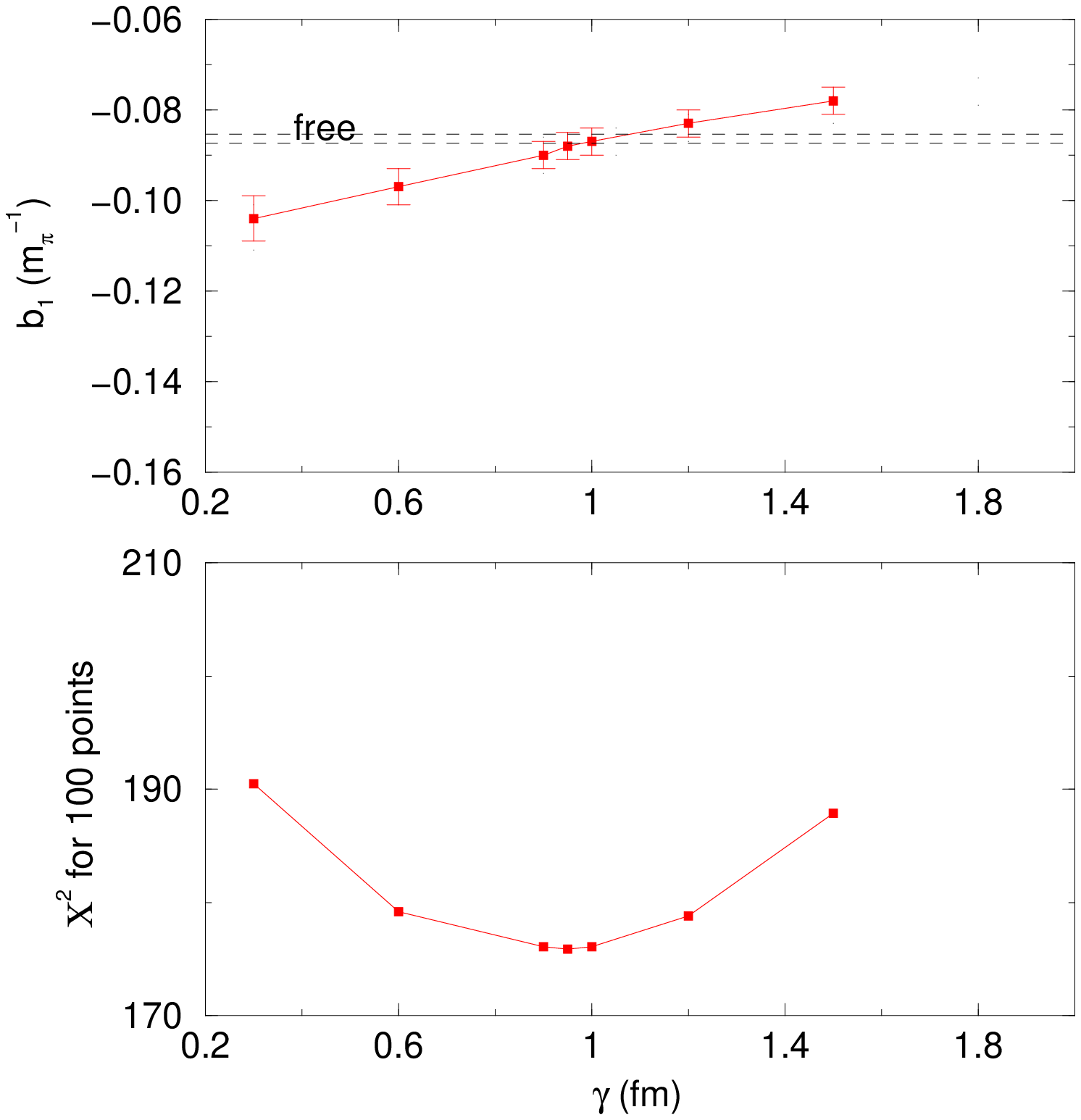} 
\caption{Fits to 100 $\pi^-$-atom data points from Ne to U as a function of 
$\gamma$. Left: density-independent (DI) $b_1$. Right: density-dependent (DD) 
$b_1$. Figure adapted from Ref.~\cite{fg07}.} 
\label{fig:pi1} 
\end{center} 
\end{figure} 

Density independent optical potential global fits to pionic atom data across 
the periodic table reveal an anomalous $s$-wave repulsion \cite{bfg97,fg07}, 
a major component of which is due to a too repulsive isovector $\pi N$ 
amplitude $b_1$ with respect to the free-space value $b_1^{\rm free}$. 
This is demonstrated on the l.h.s. of Fig.~\ref{fig:pi1} which shows results 
of global fits to pionic atom data as a function of a parameter $\gamma$, 
related to the difference between neutron and proton rms radii: 
\begin{equation} 
r_n - r_p = \gamma \frac{N-Z}{A} + \delta.  
\label{eq:gamma} 
\end{equation} 
Applying a finite-range folding with rms radius 0.9 fm to the $p$-wave part 
of $V_{\rm opt}^{\pi}$, the resulting $\chi^2$ minimum for a `skin' neutron 
distribution is obtained at $\gamma=1.1\pm 0.1$~fm.{\footnote
{For a recent discussion of the role of neutron distributions in hadronic 
atoms, see Ref.~\cite{friedman09}.}} 
The r.h.s. of the figure shows results of global fits with empirical energy 
dependence imposed on the $s$-wave amplitudes $b_0$ and $b_1$ \cite{fg04}, 
and more importantly with a DD renormalization of $b_1$: 
\begin{equation} 
b_{1}(\rho)=\frac{b_{1}}{1 - {{\sigma \rho} \over {m_{\pi}^2 f_{\pi}^2}}}\,, 
\label{eq:weise} 
\end{equation} 
where $f_{\pi}=92.4$~MeV is the pion weak decay constant and $\sigma \approx 
50$~MeV is the $\pi N$ $\sigma$ term. Eq.~(\ref{eq:weise}) was derived by 
Weise \cite{weise01} considering an in-medium extension of the 
Tomozawa-Weinberg (TW) LO chiral limit for $b_1$ in terms of 
$f_{\pi}$ \cite{tw66} which is then related to the 
quark condensate $<{\bar q}q>$: 
\begin{equation} 
b_{1}(\rho)=-\frac{\mu_{\pi N}}{8 \pi f^{2}_{\pi}(\rho)}\,,~~~~~~~~~~
\frac{f_\pi^2(\rho)}{f_\pi^2} = \frac{<\bar q q>_{\rho}}{<\bar q q>_0} 
\simeq {1 - {{\sigma \rho} \over {m_{\pi}^2 f_{\pi}^2}}}\,.  
\label{eq:gor} 
\end{equation}  
The figure makes it evident that the magnitude of $b_1$ on the r.h.s., 
following the DD renormalization, is systematically smaller than that on the 
l.h.s., and at the $\chi^2$ minimum it agrees perfectly with $b_1^{\rm free}$. 

\begin{table}[thb] 
\caption{Values of $b_1$ derived from density-independent fits to pionic atom 
data sets listed in Ref.~\cite{fg03}. For comparison, 
$b_1^{\rm free}=-0.0864\pm 0.0010~m_\pi ^{-1}$ \cite{marton07}. `Deep' refers 
to $1s$ $\pi^-$ `deeply bound' states in $^{205}$Pb \cite{geissel02} and 
$^{115,119,123}$Sn \cite{suzuki04} observed in $({\rm d},{^3{\rm He}})$ 
experiments at GSI.} 
\label{tab:errors} 
\begin{center} 
\begin{tabular}{lccc} 
\br 
data&global&light $N=Z$&`deep'\\
    &$^{20}$Ne to $^{238}$U&+~`deep' $1s$ only&$1s$ only\\
\mr
points&100&20&8\\ 
$b_1(m_\pi ^{-1})$&$-0.104\pm 0.006$&$-0.104\pm 0.013$&$-0.130\pm 0.036$\\
\br 
\end{tabular} 
\end{center} 
\end{table} 

A similar conclusion was reached in Refs.~\cite{kkw03,suzuki04} from 
measurements of $1s$ `deeply bound' pionic atoms of Sn isotopes. The advantage 
of using $1s$ `deeply bound' levels is that the $p$-wave $\pi N$ interaction 
plays there a secondary role, but this merit is more than compensated by the 
considearably increased errors associated with smaller data sets, as shown 
in Table~\ref{tab:errors}. The uncertainty listed in the fourth column makes 
it clear that the `deeply bound' atoms alone do not give conclusive evidence 
for the need to renormalize $b_1$. 
In fact, Suzuki {\it et al.} \cite{suzuki04} considered `deeply bound' $1s$ 
levels in three Sn isotopes together with `normal' $1s$ levels in $^{16}$O, 
$^{20}$Ne and $^{28}$Si, 12 data points in total yielding 
$b_1=-0.1149\pm 0.0074~m_\pi ^{-1}$. However, this small uncertainty excludes 
the uncertainty from the $p$-wave $\pi N$ potential which was held fixed in 
their analysis. A more realistic uncertainty for this type of deduction is 
given in column 3 of the table.  

\begin{figure}[hbt]  
\begin{center}
\includegraphics[width=0.45\textwidth]{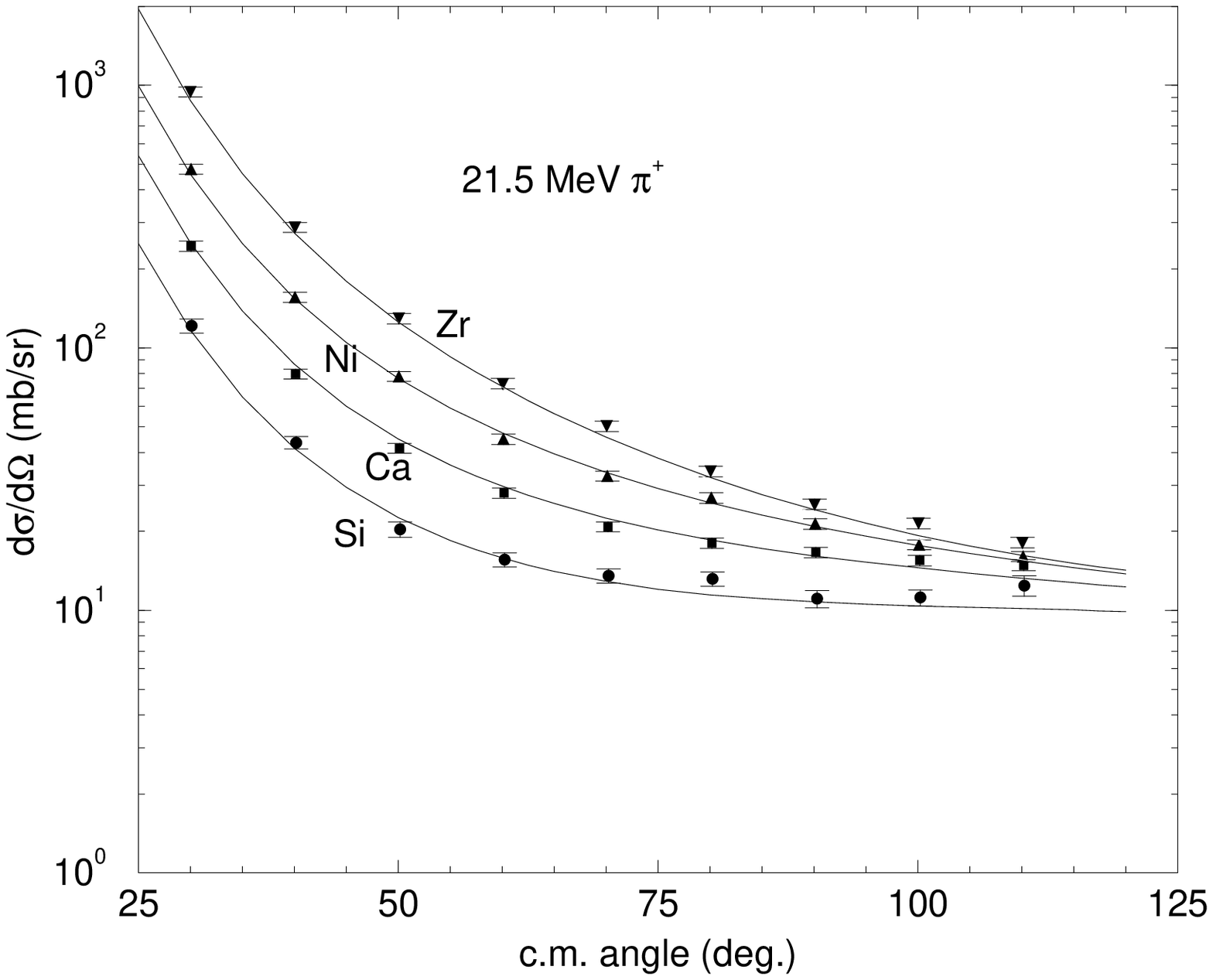} 
\hspace{3mm}
\includegraphics[width=0.45\textwidth]{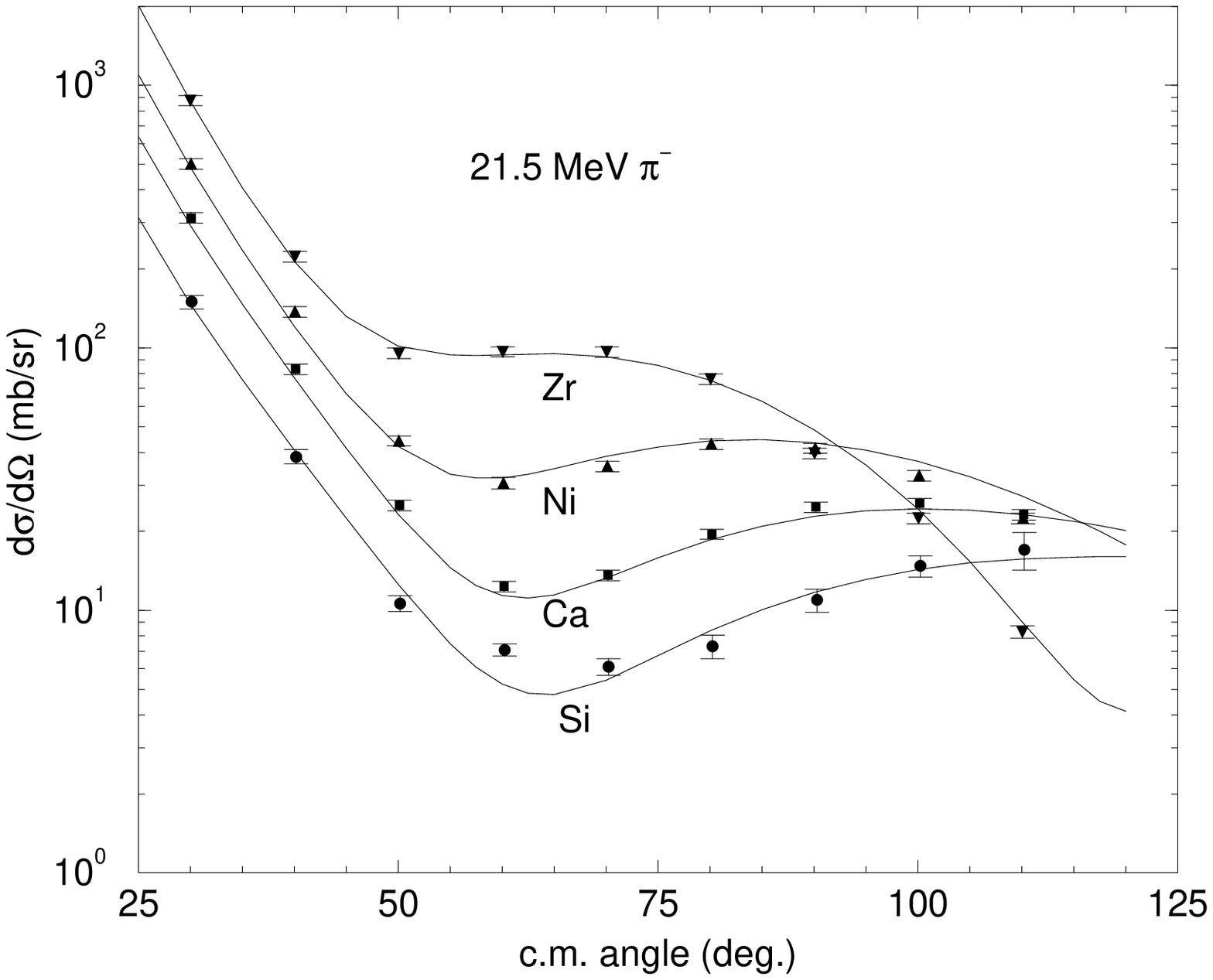} 
\caption{Low energy $\pi^{\pm}$ nucleus elastic scattering angular 
distributions \cite{friedman04} reproduced with $b_1(\rho)$ ansatz, 
Eq.~(\ref{eq:weise}).}
\label{fig:pi2} 
\end{center}
\end{figure} 

The renormalization of $b_1$ derived from pionic atoms is consistent with 
that shown recently to be also required in low energy $\pi$-nucleus 
scattering, as demonstrated in Fig.~\ref{fig:pi2} for 21.5 MeV $\pi^{\pm}$ 
scattered off isotopes of Si, Ca, Ni and Zr at PSI \cite{friedman04}.

\section{$\Sigma$ Nuclear repulsion from $\Sigma^-$ atoms} 
\label{sec:sigma} 
\vspace{0.5cm} 

\begin{figure}[t] 
\begin{center}
\includegraphics[width=0.45\textwidth]{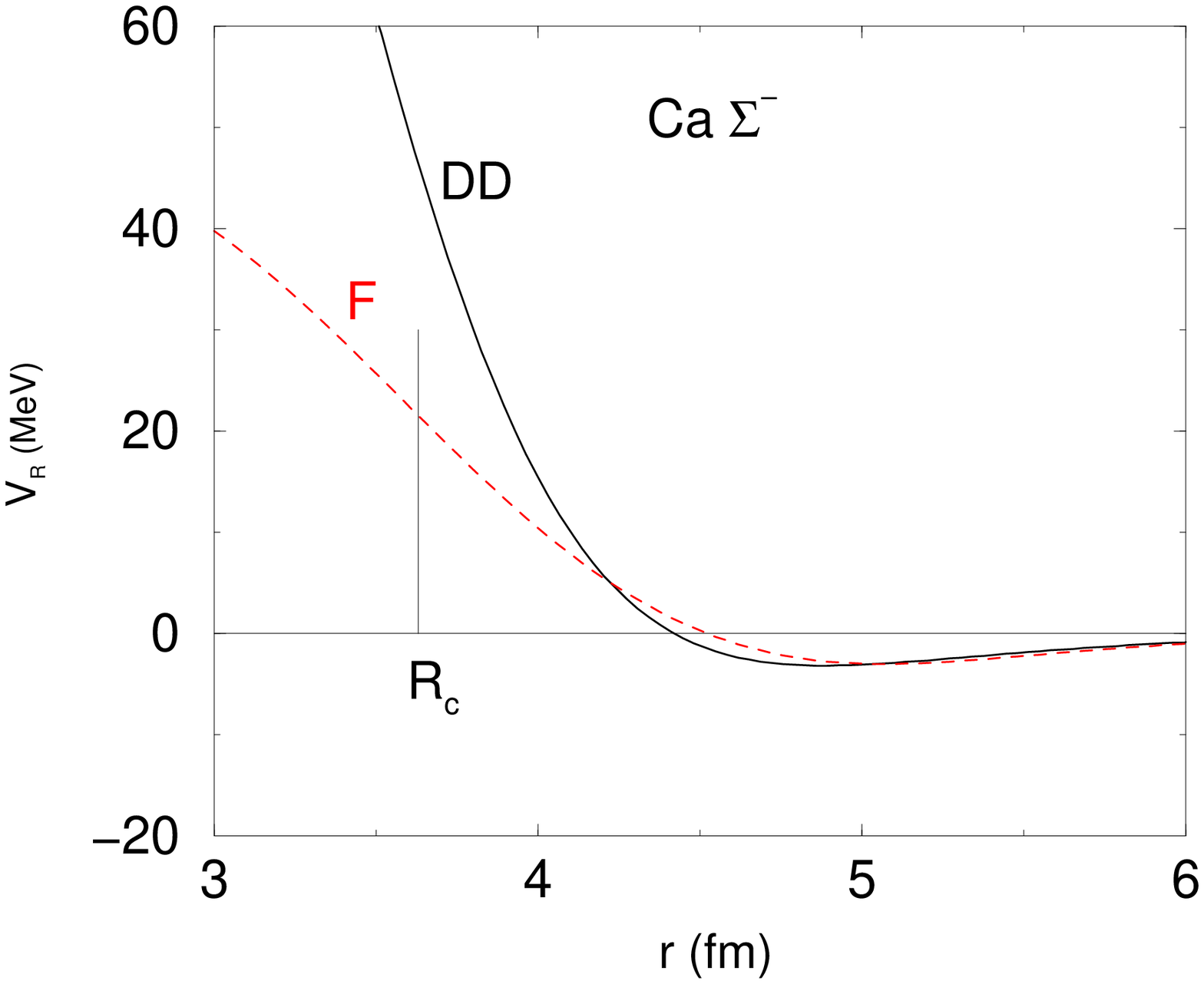} 
\hspace{3mm} 
\includegraphics[width=0.45\textwidth]{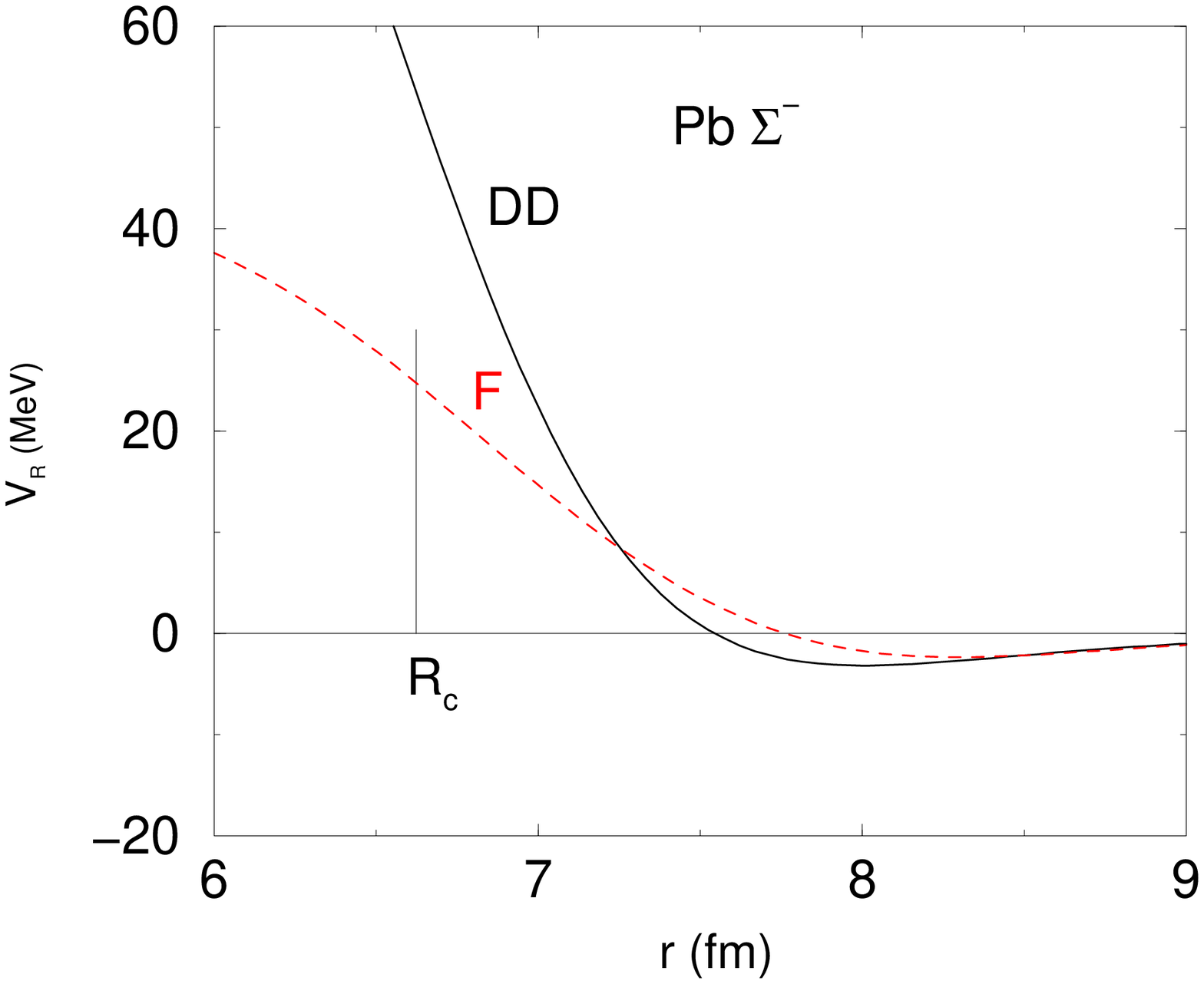} 
\caption{$V_{\rm R}^{\Sigma}$ from fits to $\Sigma^-$ atoms \cite{fg07}. 
The transition from attraction to repulsion occurs well outside of $R_c$, 
the half-density radius of the charge distribution in Ca (l.h.s.) and Pb 
(r.h.s.).} 
\label{fig:sigma1} 
\end{center} 
\end{figure}

\begin{figure}[bh] 
\begin{center}
\includegraphics[width=0.45\textwidth]{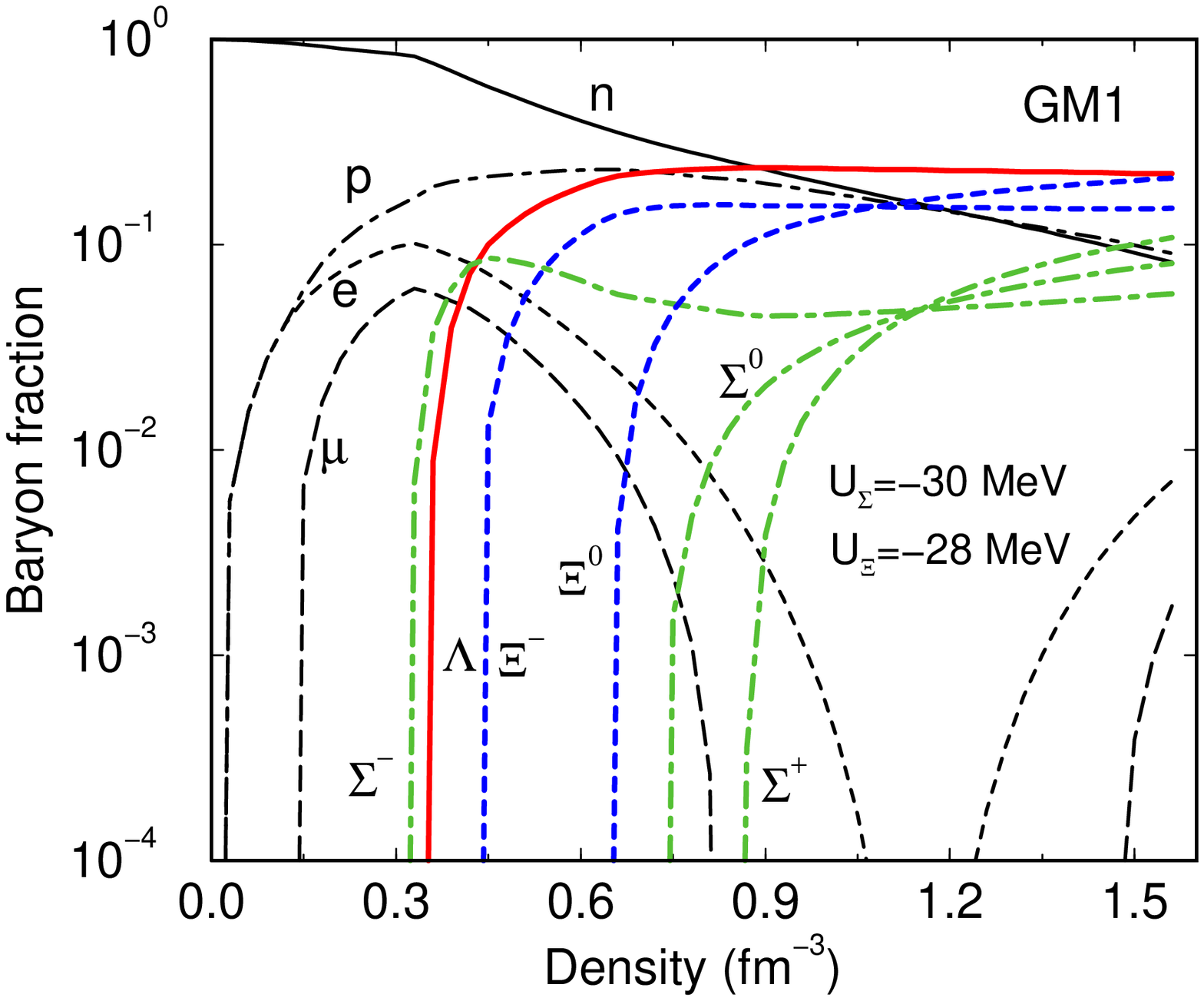} 
\hspace{3mm} 
\includegraphics[width=0.45\textwidth]{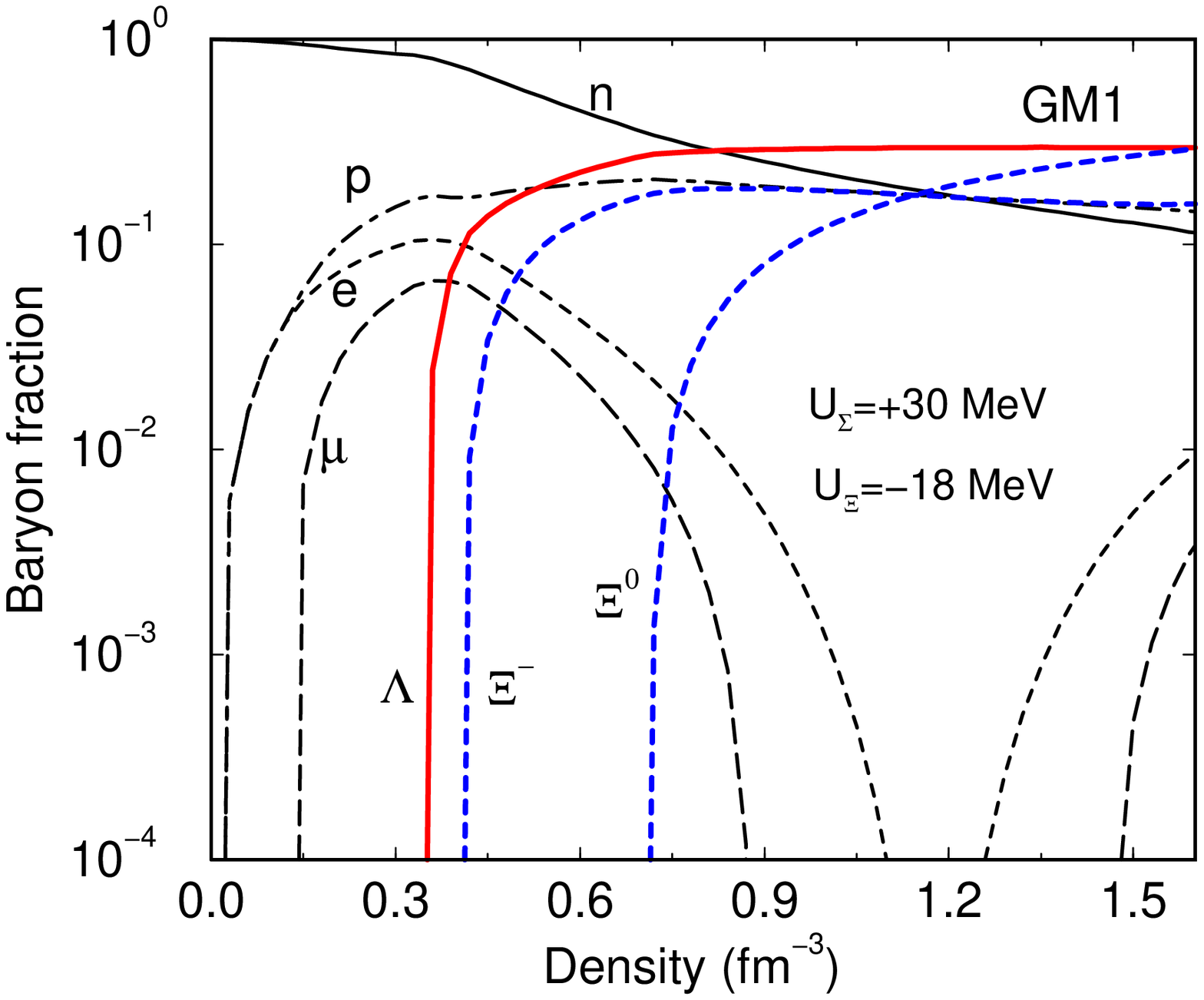} 
\caption{RMF calculations of baryon \& lepton fractions in neutron star 
matter for different scenarios of hyperon nuclear potential depths $U_Y$. 
Figure adapted from Ref.~\cite{schaffner10}.} 
\label{fig:sigma2} 
\end{center} 
\end{figure}

A vast body of $(K^-,\pi^{\pm})$ spectra indicate a repulsive and moderately 
absorptive $\Sigma$ nuclear potential $V^{\Sigma}$, with a substantial isospin 
dependence \cite{dmg89,bart99}. These data, including recent $(\pi^-,K^+)$ 
spectra \cite{noumi02} and related DWIA analyses \cite{kohno04}, provide 
credible evidence that $\Sigma$ hyperons generally do not bind in nuclei. 
A repulsive component of a DD $\Sigma$ nuclear potential was already deduced 
in the mid 1990s from $\Sigma^-$ atom data \cite{bfg94,mfgj95}, as shown in 
Fig.~\ref{fig:sigma1}. In fact, $V_{\rm R}^{\Sigma}$ is attractive at low 
densities outside the nucleus, as enforced by the observed `attractive' 
$\Sigma^-$ atomic level shifts, changing into repulsion on approach of the 
nuclear radius. The precise magnitude and shape of $V_{\rm R}^{\Sigma}$ 
within the nucleus, however, are model dependent as demonstrated by the 
difference between potentials DD and F (defined in the Appendix). This 
repulsion bears interesting consequences for the balance of strangeness in 
the inner crust of neutron stars, primarily by delaying to higher densities, 
or even aborting the appearance of $\Sigma^-$ hyperons, as shown in 
Fig.~\ref{fig:sigma2}. 

\begin{table} 
\caption{Isoscalar and isovector $\Sigma$ nucleus potentials, 
Eq.~(\ref{eq:lane}) in MeV, calculated in Nijmegen soft-core potential models 
\cite{rijken10} at $k_F=1.35~{\rm fm}^{-1}$, corresponding to nuclear-matter 
density.} 
\label{tab:sig} 
\begin{center} 
\begin{tabular}{lccccccc} 
\br 
&97f&04d&06d&08a&08b&phenom.&Ref.\\
\mr 
$V_0^{\Sigma}$&$-13.9$&$-26.0$&$-1.2$&$+13.4$&$+20.3$&$+30\pm 20$&\cite{fg07,
kohno04}\\ 
$V_1^{\Sigma}$&$-30.4$&$+30.4$&$+52.6$&$+64.5$&$+85.2$&$\approx +80$&\cite
{dgm84}\\ 
\br 
\end{tabular} 
\end{center} 
\end{table} 

\begin{figure} 
\begin{center}
\includegraphics[width=7.5cm,height=5.0cm]{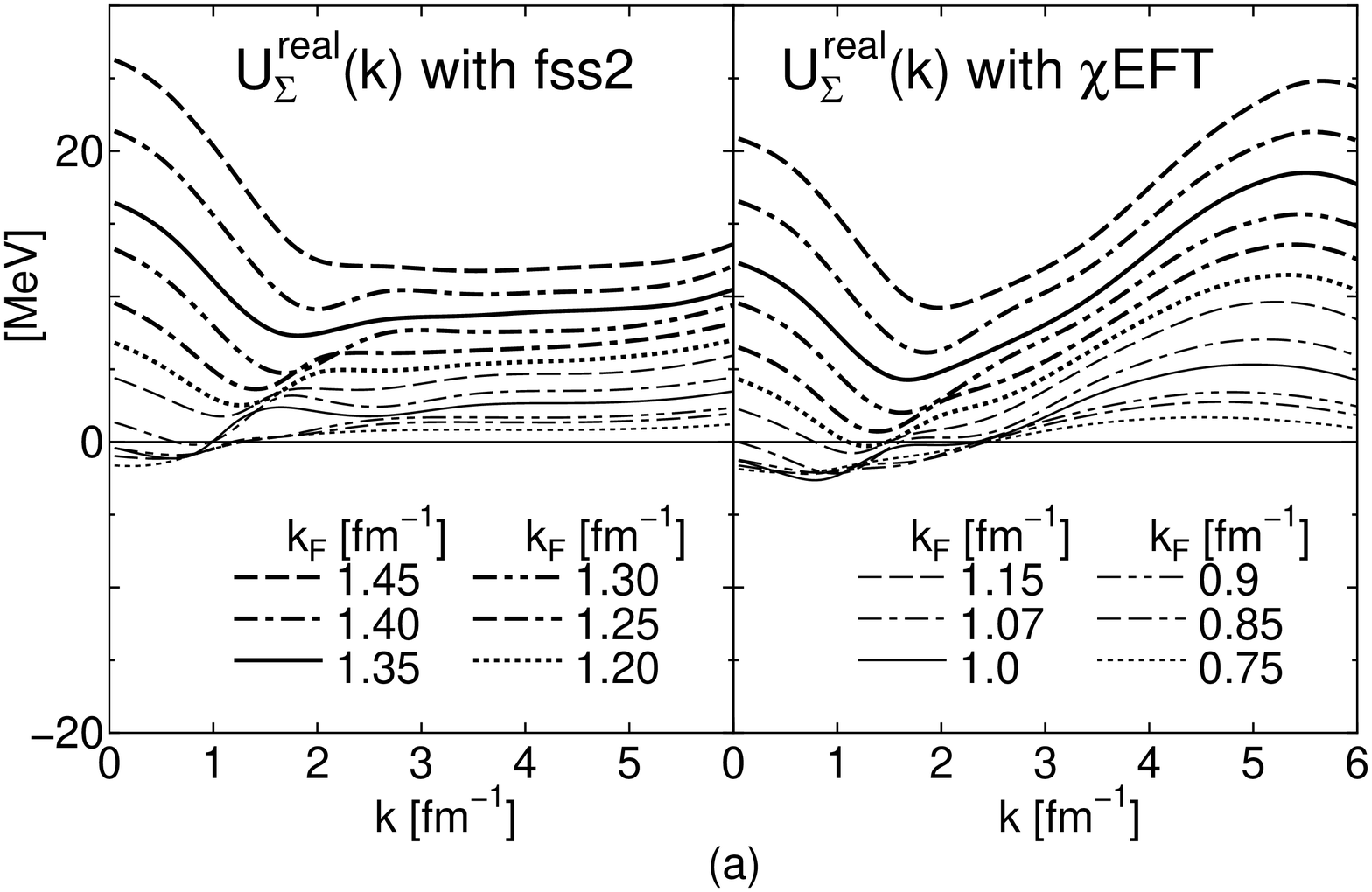} 
\hspace{3mm} 
\includegraphics[width=7.5cm,height=5.0cm]{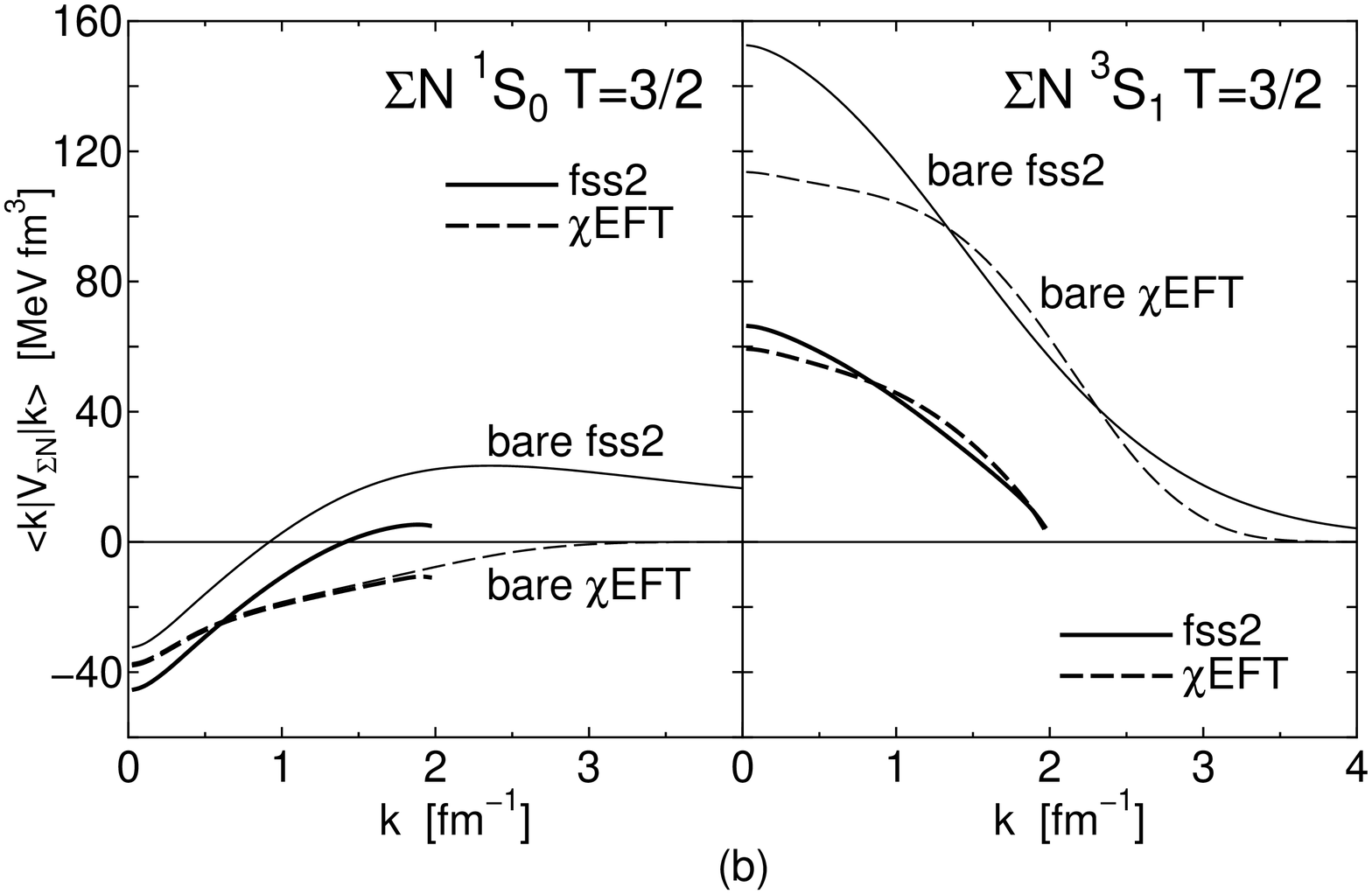} 
\caption{Left: Isoscalar $\Sigma$ nuclear potentials calculated in two 
models, Refs.~\cite{fujiwara07,polinder06}. \\ Right: $(\Sigma N)_{T=3/2}$ 
potentials calculated in these models. Figure adapted from 
Ref.~\cite{kohno10}.} 
\label{fig:sigma3} 
\end{center} 
\end{figure}

The $G$-matrices constructed from Nijmegen soft-core potential models 
have progressed throughout the years to produce $\Sigma$ repulsion in 
symmetric nuclear matter, as demonstrated in Table~\ref{tab:sig} using 
the parametrization 
\begin{equation} 
\label{eq:lane} 
V_{R}^{\Sigma} = V_0^{\Sigma} + \frac{1}{A}~V_1^{\Sigma}~{\bf T}_A{\cdot}
{\bf t}_{\Sigma} ~. 
\end{equation} 
In the latest Nijmegen ESC08 model \cite{rijken10}, this repulsion is 
dominated by repulsion in the $T=3/2,~{^3S_1}-{^3D_1}~\Sigma N$ channel where 
a strong short distance Pauli exclusion repulsion for quarks arises in SU(6) 
quark-model RGM \cite{fujiwara07} and in chiral EFT \cite{polinder06} 
calculations, as seen on the r.h.s. of Fig.~\ref{fig:sigma3}. These model 
calculations also lead to $\Sigma$ nuclear repulsion, shown in momentum space 
on the l.h.s. of the figure. A strong repulsion appears also in a recent SU(3) 
chiral perturbation calculation \cite{kaiser05} which yields 
$V_0^{\Sigma}\approx 60$~MeV. Phenomenologically $V_0^{\Sigma} > 0$ 
and $V_1^{\Sigma} > 0$, as listed in the table, and the resulting 
$\Sigma$-nuclear potential $V_{R}^{\Sigma}$ is repulsive.{\footnote
{In the case of $^4_\Sigma$He, the only known quasibound $\Sigma$ 
hypernucleus \cite{hayano89,nagae98}, the isovector term provides substantial 
attraction owing to the small value of $A$ towards binding the $T=1/2$ 
hypernuclear configuration, while the isoscalar repulsion reduces the 
quasibound level width \cite{harada98}.}} 

\section{$\bar K$-nucleus potentials from $K^-$ atoms} 
\label{sec:k} 
\vspace{0.5cm} 

The gross features of low-energy $\bar K N$ physics are encapsulated 
in the leading-order Tomozawa-Weinberg (TW) vector term of the chiral 
effective Lagrangian~\cite{tw66}. The Born approximation for the 
$\bar K$-nuclear potential $V_{\rm TW}^{\bar K}$ due to this TW 
interaction term yields a sizable attraction: 
\begin{equation} 
\label{eq:chiral} 
V_{\rm TW}^{\bar K}=-\frac{3}{8f_{\pi}^2}~\rho\sim 
-55~\frac{\rho}{\rho_0}~~~~({\rm MeV}) 
\end{equation} 
for $\rho _0 = 0.16$ fm$^{-3}$. Iterating the TW term plus the less 
significant NLO terms, within an {\it in-medium} coupled-channel approach 
constrained by the $\bar K N - \pi \Sigma - \pi \Lambda$ data near the 
$\bar K N$ threshold, roughly doubles this $\bar K$-nucleus attraction 
\cite{bnw05}. 
A major uncertainty in these chirally based studies arises from fitting 
the $\Lambda(1405)$ resonance by the imaginary part of the $(\pi\Sigma)_{I=0}$ 
amplitude calculated within the same coupled channel chiral scheme. 
Yet, irrespective of this uncertainty, the $\Lambda(1405)$ which may be 
viewed as a $K^-p$ quasibound state quickly dissolves in the nuclear medium 
at low density, so that the repulsive free-space scattering length $a_{K^-p}$, 
as function of $\rho$, becomes {\it attractive} well below $\rho _0$. 
Adding the weakly density dependent  $I=1$ attractive scattering length 
$a_{K^-n}$, the resulting in-medium $\bar K N$ isoscalar scattering length 
$b_0(\rho)={\frac{1}{2}}(a_{K^-p}(\rho)+a_{K^-n}(\rho)$) translates into 
a strongly attractive $V^{\bar K}$ \cite{cfgm01,weise08}: 
\begin{equation} 
\label{eq:trho} 
V_{R}^{\bar K}(\rho)\sim -{\frac{2\pi}{\mu_{KN}}}{\rm Re}~b_0(\rho_0)~\rho_0~
\frac{\rho}{\rho_0}\approx -110~\frac{\rho}{\rho_0}~~~({\rm MeV})\,. 
\end{equation} 
Shallower potentials, $V_{R}^{\bar K}(\rho_0)\sim -(40-60)$ MeV, 
were obtained by imposing a Watson-like self-consistency 
requirement \cite{cfgm01,ramoset00}. It turns out, however, that stronger 
attraction, $V_{R}^{\bar K}(\rho_0)\sim -(80-90)$ MeV, arises in similar 
chiral approaches \cite{cs10} when imposing the same requirement while 
considering the energy dependence of the in-medium $\bar K N$ scattering 
amplitude below threshold \cite{cfggm11}. 

\begin{figure} 
\begin{center}
\includegraphics[width=0.45\textwidth]{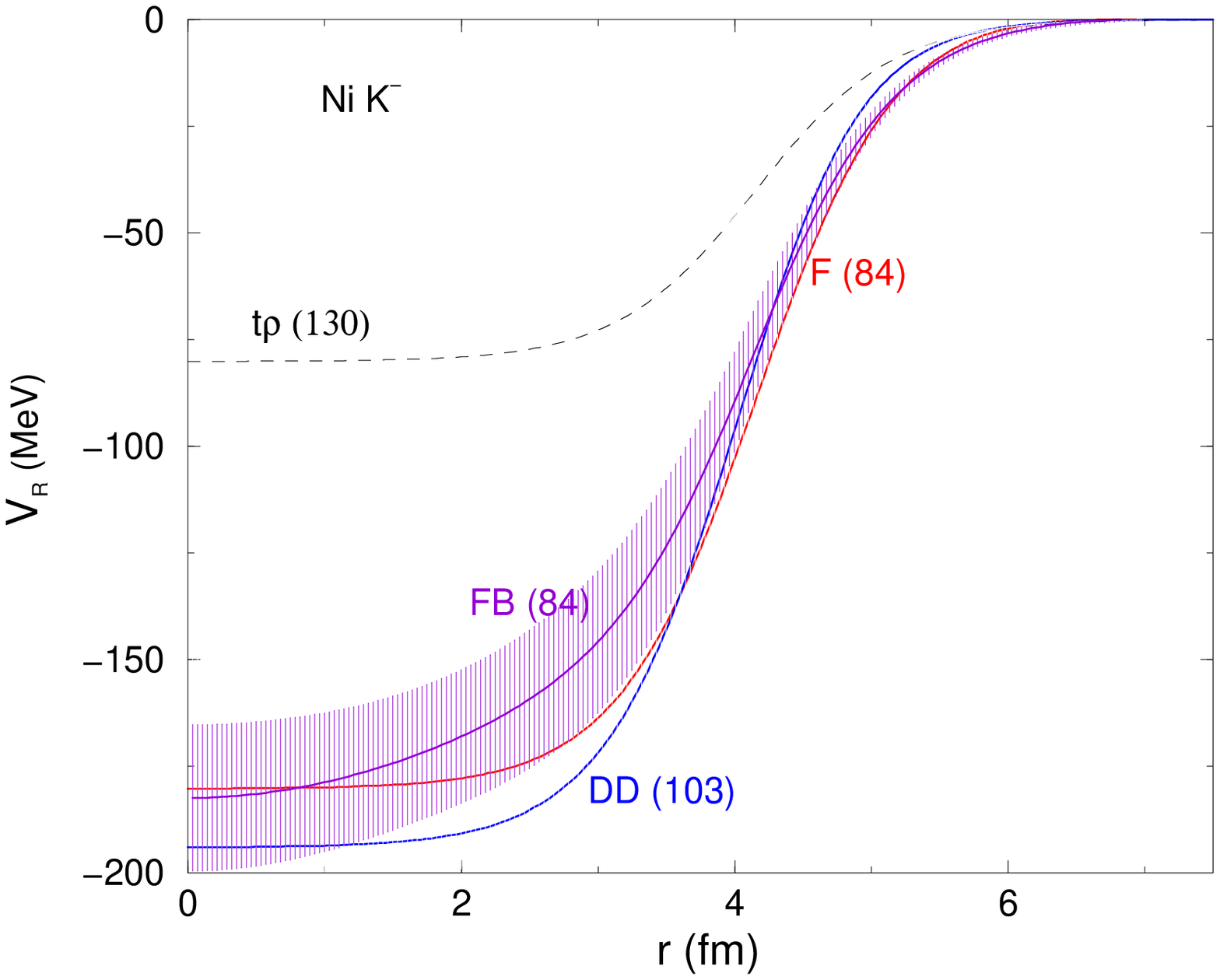} 
\hspace{3mm} 
\includegraphics[width=0.45\textwidth]{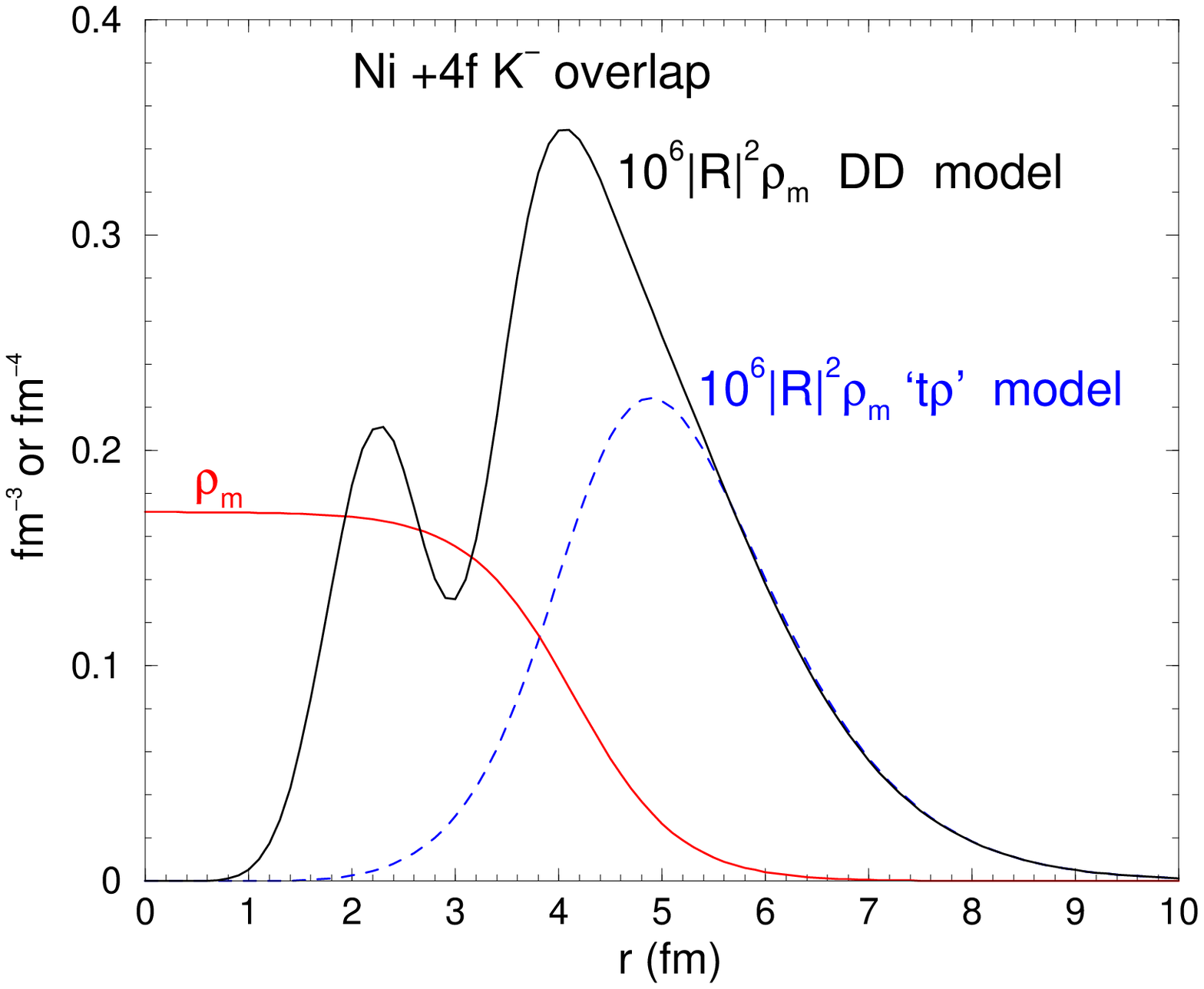} 
\caption{Left: $V_{R}^{\bar K}$ in Ni from global fits to 65 $K^-$ atom data 
points, with $\chi^2$ values \\in parentheses \cite{fg07}. The shaded area 
approximates the uncertainty in the Fourier-Bessel fit. \\ Right: Overlap of 
$K^-$ atomic $4f$ radial wavefunctions $R$ squared with Ni matter density 
$\rho_m$.} 
\label{fig:kaon1} 
\end{center} 
\end{figure}

Comprehensive fits to the strong-interaction shifts and widths of $K^-$-atom 
levels, begun in the mid 1990s \cite{fgb93}, have yielded DD deeply attractive 
and strongly absorptive optical potentials with nuclear-matter depth 
$-V_{R}^{\bar K}(\rho_0)\sim (150-200)$ MeV at threshold \cite{fgb93}. 
The l.h.s. of Fig.~\ref{fig:kaon1} illustrates for $^{58}$Ni the real part 
of $\bar K$-nucleus potentials obtained from a global fit to the data in 
several models and, in parentheses, the corresponding values of $\chi ^2$ 
for 65 $K^-$-atom data points. A model-independent Fourier-Bessel (FB) 
fit \cite{bf07} is also shown, within an error band. Just three terms in the 
FB series, added to a $t\rho $ potential, suffice to achieve a $\chi ^2$ as 
low as 84 and to make the potential extremely deep, in agreement with the 
density-dependent best-fit potentials DD and F. In particular, potential F 
provides by far the best fit ever reported for any global $K^-$-atom data 
fit~\cite{mfg06}, and the lowest $\chi ^2$ value as reached by the FB method. 

Shown on the r.h.s. of Fig.~\ref{fig:kaon1} are overlaps of the $4f$ atomic 
radial wavefunction squared with the matter density $\rho_m$ in $^{58}$Ni 
for two of the models exhibited on the l.h.s. of the figure. 
The $4f$ atomic orbit is the last circular $K^-$ 
atomic orbit from which the $K^-$ meson undergoes nuclear absorption. 
The figure demonstrates that, whereas this overlap for the shallower $t\rho$ 
potential peaks at nuclear density of order $10\%$ of $\rho_0$, it peaks at 
about $60\%$ of $\rho_0$ for the deeper DD potential and has a secondary peak 
well inside the nucleus. The double-peak structure indicates the existence of 
a $K^-$ strong-interaction $\ell=3$ quasibound state for the DD potential. 
It is clear that whereas within the $t\rho$ potential there is no sensitivity 
to the interior of the nucleus, the opposite holds for the density dependent F 
potential which accesses regions of full nuclear density. This owes partly 
to the smaller imaginary part of F. 

\begin{table}[t] 
\caption{Full and reduced $K^-$ atom data set fits. The reduced set consists 
of $2p,3d,4f,5g,7i$ shifts, widths and yields in C, Si, Ni, Sn and Pb targets, 
respectively \cite{friedman10}.}
\label{tab:reduced} 
\begin{center} 
\begin{tabular}{lllllllll} 
\br
&&\multicolumn{3}{c}{shallow potential}&&\multicolumn{3}{c}{deep potential}\\ 
N&&Re~$b(\rho_0)$&Im~$b(\rho_0)$&$\chi^2$&&Re~$b(\rho_0)$&
Im~$b(\rho_0)$&$\chi^2$\\
\mr 
65&&0.62$\pm$0.05&0.93$\pm$0.04&130&&1.44$\pm$0.03&0.59$\pm$0.03&84\\
15&&0.78$\pm$0.13&0.92$\pm$0.11&44&&1.47$\pm$0.05&0.55$\pm$0.06&26\\
\br 
\end{tabular} 
\end{center} 
\end{table} 

Given the repercussions of deeply attractive potentials on the equation 
of state of dense matter, it is important to explore the stability of 
these best-fit solutions to variations in the data selection and the 
fitting procedure. The most obvious question to ask is whether the 
resulting best-fit potentials depend strongly on the size, composition 
and accuracy of the data set studied. Regarding size and composition, 
following an earlier discussion \cite{fgmc99} it has been observed 
recently \cite{friedman10} that the DD deep potentials and the DI 
relatively shallow potentials, as well as the superiority of DD to 
DI in terms of quality of fit, persist upon decreasing the size of 
the data set. This is demonstrated in Table~\ref{tab:reduced} upon 
reducing the 65 data point global set down to 15 data points from five 
targets spread over the entire periodic table (C, Si, Ni, Sn, Pb). 
Similar results hold for any four out of these five targets. 
It makes sense then to repeat some of the $30-40$ years old $K^-$ atom 
measurements, making use of modern techniques, in order to acquire 
a minimum size canonical set of data with reduced statistical errors 
and with common systematics. For the specific set proposed in 
Ref.~\cite{friedman10}, the (lower level) widths are directly measurable 
yet not excessively large to make it difficult to observe the feeding 
X-ray transition above the background. Similarly, the relative yields 
of the upper to lower level transitions are of the order of $10\%$ and 
higher. Fitting to such a data set with improved accuracy could resolve 
the issue of deep vs. shallow potentials and determine how deep is `deep'. 

\begin{figure}[hbt]  
\begin{center}
\includegraphics[width=0.45\textwidth]{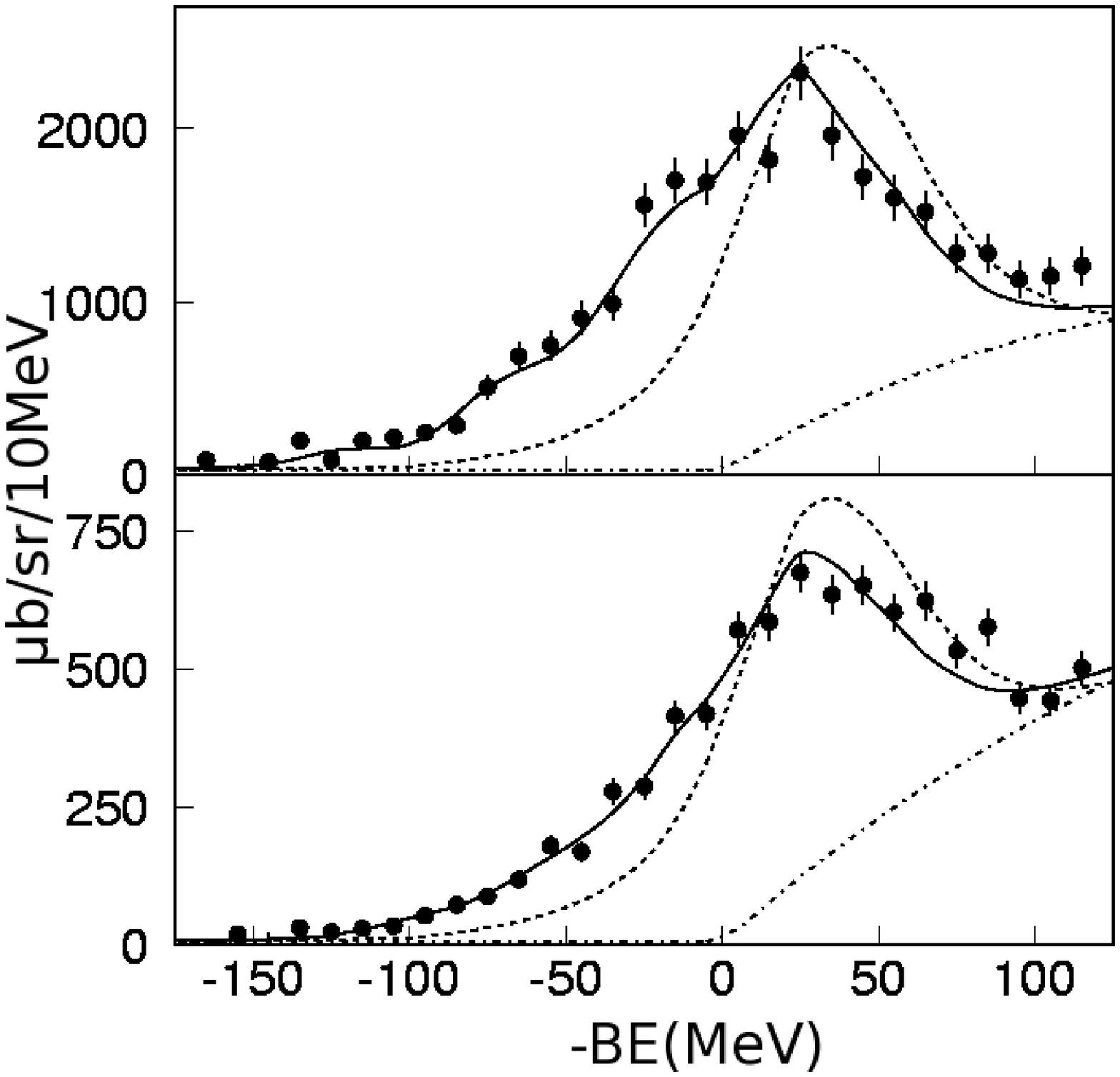} 
\hspace{3mm} 
\includegraphics[width=0.5\textwidth]{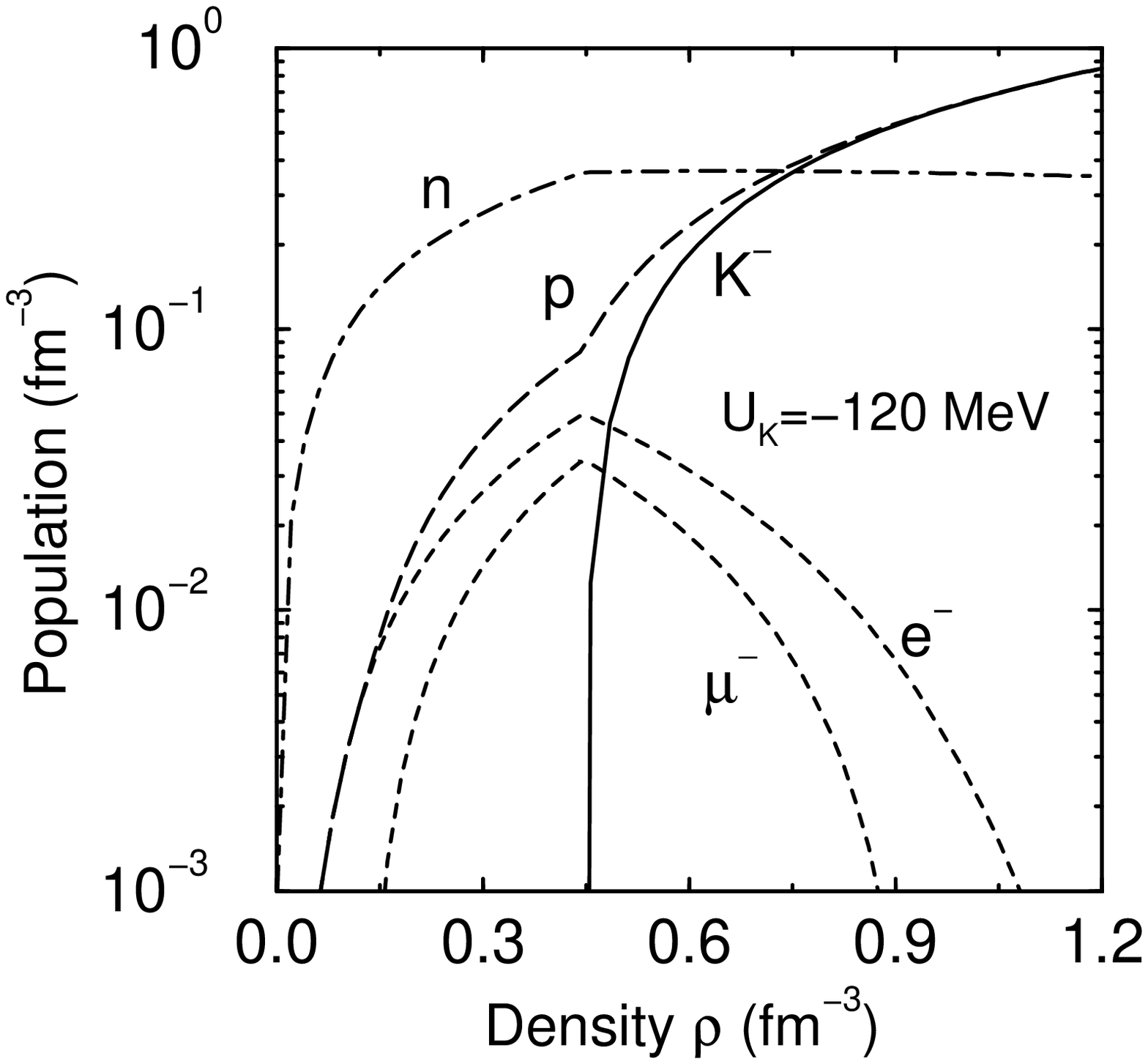} 
\caption{Left: KEK-PS E548 missing mass $(K^-,n)$ (upper) \& $(K^-,p)$ (lower) 
spectra on $^{12}$C at $p_{K^-}=1$~GeV/c \cite{kish07}. Right: calculated 
neutron-star population as a function of density. The neutron density stays 
nearly constant once kaons condense \cite{gsb99}.} 
\label{fig:kaon2} 
\end{center} 
\end{figure} 
 
A fairly new and independent evidence in favor of extremely deep 
$\bar K$-nucleus potentials is provided by $(K^-,n)$ and $(K^-,p)$ 
spectra taken at KEK on $^{12}$C \cite{kish07} and very recently 
also on $^{16}$O \cite{kish09} at $p_{K^-}=1$ GeV/c. The $^{12}$C 
spectra are shown on the l.h.s. of Fig.~\ref{fig:kaon2}, where the solid lines 
represent calculations (outlined in Ref.~\cite{yamagata06}) using potential 
depths in the range $160-190$ MeV. The dashed lines correspond to using 
relatively shallow potentials of depth about 60 MeV which may be considered 
excluded by these data. However, Magas {\it et al.}~\cite{magas10} have 
recently expressed concerns about protons of reactions other than those 
{\it directly} emanating in the $(K^-,p)$ reaction and which could explain 
part of the bound-state region of the measured spectrum without invoking 
a very deep $\bar K$-nuclear potential. A sufficientlly deep potential 
would allow quasibound states bound by over 100 MeV, for which the major 
$\bar K N \to \pi \Sigma$ decay channel is blocked, resulting in relatively 
narrow $\bar K$-nuclear states. Of course, a fairly sizable extrapolation 
is involved in this case using an energy-independent potential determined 
largely near threshold. Furthermore, the best-fit $V_{I}^{\bar K}$ imaginary 
depths of $40-50$ MeV imply that $\bar K$-nuclear quasibound states are broad, 
as studied in Refs.~\cite{mfg06,gazda07}.

\begin{figure}[hbt]
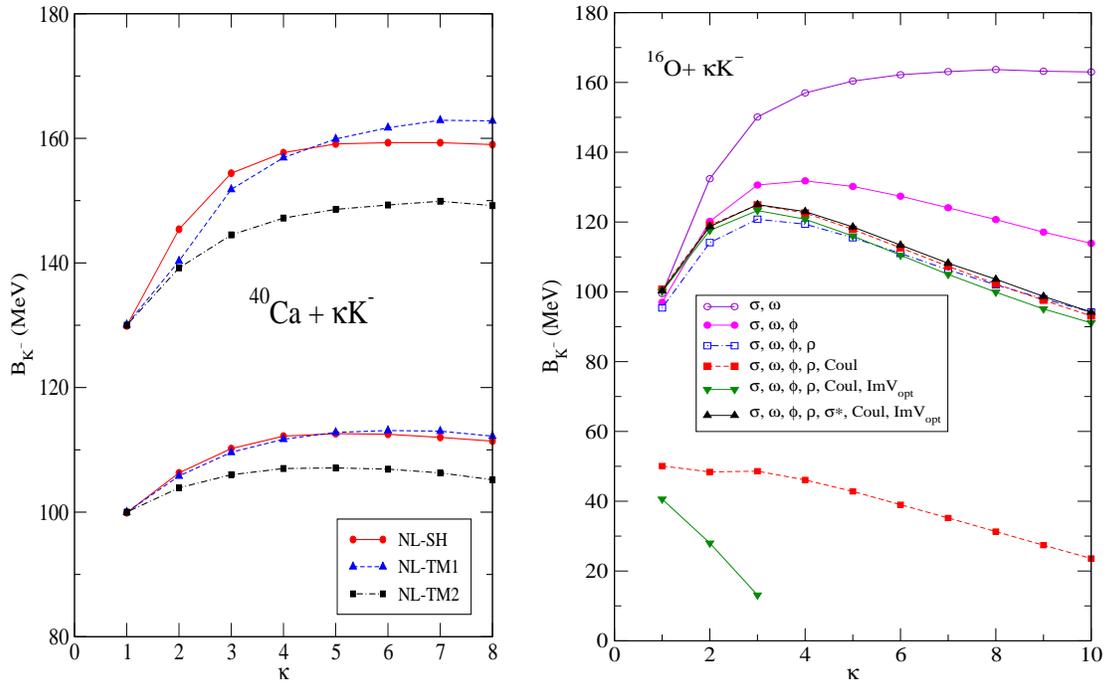
 
\begin{center}
\includegraphics[width=6.5cm,height=9.0cm]{multik40ca.eps} 
\hspace{3mm} 
\includegraphics[width=7.5cm,height=9.0cm]{multik16O.eps} 
\caption{Saturation of $K^-$ separation energy $B_{\bar K}$ in RMF 
calculations of multi $K^-$ nuclei as a function of the number $\kappa$ of 
$K^-$ mesons \cite{gazda08}. Left: RMF model dependence. Right: dependence on 
RMF ingredients.}  
\label{fig:kaon3} 
\end{center} 
\end{figure} 

A robust consequence of the sizable $\bar K$-nucleus attraction is that 
$K^-$ condensation, when hyperon degrees of freedom are ignored, could 
occur in neutron star matter at about 3 times nuclear matter density, 
as shown on the r.h.s. of Fig.~\ref{fig:kaon2}. Comparing it with 
Fig.~\ref{fig:sigma2} for neutron stars, but where strangeness 
materialized through hyperons, one may ask whether $\bar K$ mesons 
condense also in the presence of hyperons. This question was posed 
within RMF calculations of neutron star matter long ago and answered 
negatively \cite{knorren95,schaffner96}, but only recently it was posed 
for strange hadronic matter in Ref.~\cite{gazda08} by calculating 
multi-$\bar K$ nuclear configurations. Fig.~\ref{fig:kaon3} demonstrates 
a remarkable saturation of $K^-$ separation energies $B_{K^-}$ calculated 
in multi-$K^-$ nuclei, independently of the applied RMF model as shown on 
the l.h.s. for three different nuclear RMF schemes. The r.h.s. of the 
figure demonstrates that this saturation persists already in the most 
straightforward $\sigma + \omega$ model, primarily owing to the repulsion 
induced by the vector $\omega$ field between like $\bar K$ mesons. 
The additional vector fields $\rho$ and $\phi$ only add repulsion, 
thus strengthening the saturation. The effect of $V_{I}^{\bar K}$ 
is noticeable only below $B_{\bar K}\approx 100$~MeV, as seen by the 
departure of the lowest green line with respect to the lowest red line. 

The saturation values of $B_{K^-}$ do not allow conversion of hyperons 
to $\bar K$ mesons through the strong decays $\Lambda \to p + K^-$ or 
$\Xi^-\to\Lambda +K^-$ in multi-strange hypernuclei, which therefore remain 
the lowest-energy configuration for multi-strange systems \cite{sbg00}. This 
provides a powerful argument against $\bar K$ condensation in the laboratory, 
under strong-interaction equilibrium conditions \cite{gazda08}. It does not 
apply to kaon condensation in neutron stars, where equilibrium configurations 
are determined by weak-interaction conditions. This work has been recently 
generalized to multi-$K^-$ {\it hypernuclei} \cite{gazda09}.

\appendix 
\section*{Appendix: density dependent optical potentials} 
\setcounter {section}{1} 
\vspace{0.5cm} 

Here we specify the functional form of two density dependent optical 
potentials used in studies of hadronic atoms. For a recent application 
to $K^-$ atoms, see Table 1 of Ref.~\cite{mfg06}. 

\begin{itemize} 

\item The DD form is based on modifying the effective scattering length $b_0$ 
(e.g. Eq.~(\ref{eq:trho})): 
\begin{equation} 
\label{equ:DD1} 
V^{h}(r)\sim -{\frac{2\pi}{\mu_{hN}}}~b_0~\rho(r)~~\Rightarrow~~
b_0\rightarrow b_0~+~B_0~\{\frac{\rho(r)}{\rho _0}\}^\alpha~,~~\alpha >0~, 
\end{equation} 
where $\rho _0 = 0.16~{\rm fm}^{-3}$ is a central nuclear density. 
It is possible then to respect the `low density limit' by keeping $b_0$ fixed, 
$b_0=b_0^{\rm free}$, while varying the parameters $B_0$ and $\alpha$.
\\ 
\item The F form is based on modifying $b_0$ as follows: 
\begin{equation} 
\label{eq:DDF} 
b_0~\rightarrow ~B_0~F(r)~+~b_0~[1~-~F(r)]~~. 
\end{equation} 
The density-like function $F(r)$ is defined as 
\begin{equation} 
\label{eq:F} 
F(r)~=~\frac{1}{e^x +1}~, ~~~~x~=~\frac{r-R_x}{a_x}~.  
\end{equation} 
Clearly, $F(r)\rightarrow 1$ for $r << R_x$ which defines an internal region 
and similarly $[1-F(r)]\rightarrow 1$ for $r >> R_x$ which defines an external 
region. Thus $R_x$ forms an approximate border between internal and external 
regions, and {\it if} $R_x$ is close to the nuclear surface then the two 
regions do correspond to the high-density and low-density regions of nuclei, 
respectively. In global fits across the periodic table, $R_x$ is parametrized 
as $R_x = R_{x0}A^{1/3}+\delta_x$ and the parameters $B_0,~ R_{x0}$ and 
$\delta _x$ are varied upon in the least-squares fit, while gridding on values 
of $a_x$ around $0.5$~fm. The  parameter $b_0$ may be held fixed at its free 
$hN$ value, but the results often depend very little on its precise value.  

\end{itemize}

\section*{Acknowledgments}
\vspace{0.5cm}

On the occasion of Gerry Brown's 85th birthday Festschrift, we dedicate this 
mini review to him who commissioned our two past reviews \cite{bfg97,fg07} on 
similar subjects. This work was supported in part by the SPHERE collaboration 
within the HadronPhysics2 Project No. 227431 of the EU initiative FP7.


\section*{References}
\vspace{0.5cm}
\begin{thebibliography}{99}

\bibitem{bfg97} Batty C J, Friedman E and Gal A 1997 {\it Phys.\ Rept.} 
{\bf 287} 385 

\bibitem{fg07} Friedman E and Gal A 2007 {\it Phys.\ Rept.} {\bf 452} 89, 
and references therein  

\bibitem{friedman09} Friedman E 2009 {\it Hyp.\ Int.} {\bf 193} 33, 
and references therein 

\bibitem{fg04} Friedman E and Gal A 2004 {\it Phys.\ Lett.} B {\bf 578} 85 

\bibitem{weise01} Weise W 2001 {\it Nucl.\ Phys.} A {\bf 690} 98c 

\bibitem{tw66} Tomozawa Y 1966 {\it Nuovo\ Cimento} A {\bf 46} 707, 
Weinberg S 1966 {\it Phys.\ Rev.\ Lett.} {\bf 17} 616 

\bibitem{kkw03} Kolomeitsev E E, Kaiser N and Weise W 2003 
{\it Phys.\ Rev.\ Lett.} {\bf 90} 092501

\bibitem{suzuki04} Suzuki K {\it et al.} 2004 {\it Phys.\ Rev.\ Lett.} 
{\bf 92} 072302 

\bibitem{fg03} Friedman E and Gal A 2003 {\it Nucl.\ Phys.} A {\bf 724} 143 

\bibitem{marton07} Marton J 2007 {\it Nucl.\ Phys.} A {\bf 790} 328c  

\bibitem{geissel02} Geissel H {\it et al.} 2002 {\it Phys.\ Rev.\ Lett.} 
{\bf 88} 122301 

\bibitem{friedman04} Friedman E {\it et al.} 2004 {\it Phys.\ Rev.\ Lett.} 
{\bf 93} 122302, 2005 {\it Phys.\ Rev.} C {\bf 72} 034609 

\bibitem{dmg89} Dover C B, Millener D J and Gal A 1989 {\it Phys.\ Rept.} 
{\bf 184} 1, and references therein 

\bibitem{bart99} Bart S {\it et al.} [BNL E887] 1999 {\it Phys.\ Rev.\ Lett.} 
{\bf 83} 5238

\bibitem{noumi02} Noumi H {\it et al.} [KEK E438] 2002 
{\it Phys.\ Rev.\ Lett.} {\bf 89} 072301, 2003 {\it Phys.\ Rev.\ Lett.} 
{\bf 90} 049902(E), \\ Saha P K {\it et al.} 2004 {\it Phys.\ Rev.} C {\bf 70} 
044613 

\bibitem{kohno04} Kohno M, Fujiwara Y, Watanabe Y, Ogata K and Kawai M 2004 
{\it Prog.\ Theor.\ Phys.} {\bf 112} 895, \\ 2006 {\it Phys.\ Rev.} C {\bf 74} 
064613, Harada T and Hirabayashi Y 2005 {\it Nucl.\ Phys.} A {\bf 759} 
143, 2006 {\bf 767} 206 

\bibitem{bfg94} Batty C J, Friedman E and Gal A 1994 {\it Phys.\ Lett.} B 
{\bf 335} 273, {\it Prog.\ Theor.\ Phys.\ Suppl.} {\bf 117} 227 

\bibitem{mfgj95} Mare\v{s} J, Friedman E, Gal A and Jennings B K 1995 
{\it Nucl.\ Phys.} A {\bf 594} 311 

\bibitem{schaffner10} 
Schaffner-Bielich J 2010 {\it Nucl.\ Phys.} A {\bf 835} 279, and references 
therein 

\bibitem{rijken10} Rijken Th A, Nagels M M and Yamamoto Y 2010 
{\it Nucl.\ Phys.} A {\bf 835} 160, and references therein 

\bibitem{dgm84} Dover C B, Gal A and Millener D J 1984 {\it Phys.\ Lett.} B 
{\bf 138} 337 

\bibitem{fujiwara07} Fujiwara Y, Suzuki Y and Nakamoto C 2007 
{\it Prog.\ Part.\ Nucl.\ Phys.} {\bf 58} 439, and references therein  

\bibitem{polinder06} Polinder H, Haidenbauer J and Mei{\ss}ner U G 2006
{\it Nucl.\ Phys.} A {\bf 779} 244

\bibitem{kohno10} Kohno M 2010 {\it Phys.\ Rev.} C {\bf 81} 014003

\bibitem{kaiser05} Kaiser N 2005 {\it Phys.\ Rev.} C {\bf 71} 068201 

\bibitem{hayano89} Hayano R S {\it et al.} 1989 {\it Phys.\ Lett.} 
B {\bf 231} 355 

\bibitem{nagae98} Nagae T {\it et al.} [BNL E905] 1998 {\it Phys.\ Rev. Lett.} 
{\bf 80} 1605 

\bibitem{harada98} Harada T 1998 {\it Phys.\ Rev.\ Lett.} {\bf 81} 5287 

\bibitem{bnw05} Borasoy B, Ni{\ss}ler R and Weise W 2005 {\it Eur.\ Phys.\ J.} 
A {\bf 25} 79 

\bibitem{cfgm01} Ciepl\'{y} A, Friedman E, Gal A and Mare\v{s} J 2001 
{\it Nucl.\ Phys.} A {\bf 696} 173 

\bibitem{weise08} Weise W and H\"{a}rtle R 2008 {\it Nucl.\ Phys.} A {\bf 804} 
173 

\bibitem{ramoset00} Ramos A and Oset E 2000 {\it Nucl.\ Phys.} A {\bf 671} 
481 

\bibitem{cs10} Ciepl\'{y} A and Smejkal J 2010 {\it Eur.\ Phys.\ J.} A 
{\bf 43} 191 

\bibitem{cfggm11} Ciepl\'{y} A, Friedman E, Gal A, Gazda D and Mare\v{s} J 
2011 {\it Phys.\ Lett.} B {\bf 702} 402, see also Ciepl\'{y} A, Friedman E, 
Gal A and Krej\v{c}i\v{r}\'{i}k V 2011 {\it Phys.\ Lett.} B {\bf 698} 226 

\bibitem{fgb93} Friedman E, Gal A and Batty C J 1993 {\it Phys.\ Lett.} B 
{\bf 308} 6, 1994 {\it Nucl.\ Phys.} A {\bf 579} 518 

\bibitem{bf07} Barnea N and Friedman E 2007 {\it Phys.\ Rev.} C {\bf 75} 
022202(R) 

\bibitem{mfg06} Mare\v{s} J, Friedman E and Gal A 2006 {\it Nucl.\ Phys.} A 
{\bf 770} 84 

\bibitem{fgmc99} Friedman E, Gal A, Mare\v{s} J and Ciepl\'{y} A 1999 
{\it Phys.\ Rev.} C {\bf 60} 024314 

\bibitem{friedman10} 
Friedman E 2011 {\it Int.\ J.\ Mod.\ Phys.} A {\bf 26} 468 (Proc. Int. Conf. 
on Meson Physics, Krakow 2010) 

\bibitem{kish07} Kishimoto T 2007 {\it et al.} [KEK E548] {\it Prog.\ Theor.\ 
Phys.} {\bf 118} 181 

\bibitem{kish09} Kishimoto T 2009 {\it Nucl.\ Phys.} A {\bf 827} 321c 

\bibitem{gsb99}
Glendenning N K and Schaffner-Bielich J 1999 {\it Phys.\ Rev.} C {\bf 60} 
025803 

\bibitem{yamagata06} Yamagata J, Nagahiro H and Hirenzaki S 2006 
{\it Phys.\ Rev.} C {\bf 74} 014604 

\bibitem{magas10} Magas V K, Yamagata-Sekihara J, Hirenzaki S, Oset E and 
Ramos A 2010 {\it Phys.\ Rev.} C {\bf 81} 024609 

\bibitem{knorren95} Knorren R, Prakash M and Ellis P J 1995 {\it Phys.\ Rev.} 
C {\bf 52} 3470 

\bibitem{schaffner96} Schaffner J and Mishustin I N 1996 {\it Phys.\ Rev.} 
C {\bf 53} 1416 

\bibitem{gazda07} Gazda D, Friedman E, Gal A and Mare{\v{s}} J 2007 
{\it Phys.\ Rev.} C {\bf 76} 055204 

\bibitem{gazda08} Gazda D, Friedman E, Gal A and Mare{\v{s}} J 2008 
{\it Phys.\ Rev.} C {\bf 77} 045206 

\bibitem{sbg00} Schaffner-Bielich J and Gal A 2000 {\it Phys.\ Rev.} C 
{\bf 62} 034311, and references therein 

\bibitem{gazda09} Gazda D, Friedman E, Gal A and Mare{\v{s}} J 2009 
{\it Phys.\ Rev.} C {\bf 80} 035205 

\end{thebibliography}
\end{document}